# DynaMate2: runtime registration of expert-defined tools for agentic scientific workflow automation

*Orlando A. Mendible-Barreto[1], Ajay Vallabh[2], Ubaldo M. Córdova-Figueroa[2], and Yamil J. Colón[1]**

[1]Department of Chemical and Biomolecular Engineering, University of Notre Dame, Notre Dame, IN, 46556

[2]Department of Chemical Engineering, Universidad de Puerto Rico – Mayagüez, PR, 00680

**Corresponding author: ycolon@nd.edu*

## Abstract

Agentic large-language-model systems can coordinate scientific tools, but many implementations remain difficult for domain scientists to extend without modifying the source orchestration code or relying on unconstrained code generation. DynaMate2 is a LangGraph-based multi-agent framework for converting expert-defined Python functions into persistent AI-callable tools. The architecture separates domain execution from LLM supervision: registered tools perform scientific operations, while a supervisor LLM decomposes goals, selects specialist agents, routes inputs, and propagates outputs across steps. DynaMate2 supports: runtime tool registration from inline code, source files, and explicitly requested natural-language specifications; persistent storage of tools, agents, and conversation state; and a web interface for interactive workflow assembly. We demonstrate the framework on a molecular simulation workflow in which a single instruction retrieves a MACE foundation model, builds a NaCl-water configuration, runs an ASE molecular dynamics trajectory, and generates energy and temperature diagnostics. The demonstration illustrates how validated workflow components can be composed into a supervised agentic pipeline without rewriting the framework. DynaMate2 therefore provides a reusable template for extending LLM-based automation to research groups with existing Python workflows, while preserving the need for explicit tool validation, reproducibility logs, and deployment-specific safeguards.

# 1. Introduction

Large language model (LLM)-based agentic frameworks have rapidly evolved from simple chatbots into complex orchestrators capable of planning, executing, and analyzing results from multi-step scientific workflows.[1] In the context of computational chemistry and molecular dynamics (MD) simulations, frameworks such as ChemCrow[2], CACTUS[3], QUASAR[4], El Agente[5], MDCrow[6], and others[7–9] have demonstrated that LLMs augmented with expert-designed tools can automate tasks related to molecular design, reaction planning, and simulation setup by translating natural language into executable actions. In addition, machine learning interatomic potentials (MLIPs) have become essential tools for running such simulations at near-DFT accuracy.[10] Specialized agents such as ChatMOF[11] and other chemistry-focused frameworks[12–14] have further used retrieval-augmented generation[15] (RAG) to integrate knowledge bases, including peer-reviewed literature and simulation data, to provide context-aware suggestions that improve the reliability and consistency of MD and cheminformatics workflows.

Despite this progress, a fundamental limitation persists: existing agentic frameworks are developed for a specific, predefined series of tasks. Once built for a particular workflow, they can help users automate that pipeline.[7] However, when users need to integrate new tools, adapt to new domains, or extend the framework beyond its original scope, usability becomes severely limited. New and experienced users alike face significant technical barriers when trying to incorporate their own tools into an agentic framework, and many ultimately opt out of the approach entirely.[16,17] A remaining practical barrier is not simply whether an LLM can call scientific tools, but how a domain scientist can promote an existing, validated Python function into a persistent, inspectable, reusable tool within a multi-agent workflow without editing the orchestration code. Recent autonomous systems increasingly support tool use, planning, and even tool generation, but these capabilities are often delivered through developer-curated toolsets, task-specific agent designs, or code-generation loops whose outputs must be validated before scientific use.[18,19] This leaves a distinct unresolved problem: a lightweight protocol by which research groups can register their own established workflow components, assign them to specialist agents, preserve them across sessions, and compose them through supervised planning while keeping the scientific domain logic under user control.

The first version of DynaMate[20] provided a reusable template for automating these expert-validated workflows and custom tasks in scientific workflows. It comprised multiple MD-focused agents but effectively behaved as a single-agent system. For a given prompt, only one agent was selected to invoke one of its tools and return a response, with no inter-agent communication or coordination. This required users to manually decompose workflows into multiple prompts and iteratively steer the system. As the number of tools, agents, and targeted workflows increased, this approach became increasingly inefficient, highlighting the need for hierarchical mechanisms to enable greater autonomy and robustness.

Here we present DynaMate2, a persistent multi-agent framework designed to address this runtime extensibility problem. The framework treats tools and agents as user-managed resources: Python functions can be registered from inline code or source files, optionally generated from an explicit natural-language specification, assigned to specialist agents, and restored automatically in later sessions. The supervisor LLM is used for decomposition, routing, and context propagation, whereas scientific execution is delegated to registered functions whose source remains inspectable and editable by the user. We validate each registration pathway and demonstrate the framework on an MLIP-based molecular simulation workflow.

DynaMate2 achieves this through a hierarchical architecture built on LangGraph[21], in which a supervisor agent coordinates a pool of specialist agents, each equipped with user-registered tools. Tools can be added at runtime from inline code, existing source files, or plain-language descriptions, and all extensions persist across sessions. This approach enables researchers to build a custom agentic workflow iteratively over time, reusing and extending the same infrastructure across research projects. The framework is released as an open-source reference implementation intended to serve as a community template, with guidelines for extending the tool library and agent pool to arbitrary scientific domains.

The structure of this paper is as follows: Section 2 describes the hierarchical architecture of the agentic framework. Section 3 presents the Tool Registration Protocol, a step-by-step guide for integrating expert-defined code into the framework. Section 4 demonstrates the full capabilities of the system through an end-to-end molecular dynamics workflow. Section 5 discusses current limitations and future directions.

## 2. Dynamic Hierarchical Architecture

Unlike LangChain-style linear chains or shallow-branching pipelines, where components are composed in fixed sequences with limited feedback and looping, LangGraph treats the agent system as a dynamic state machine with nodes and edges.[21,22] The nodes are agents and tools that can iteratively call one another, maintain memory, and respond to evolving conversation states. LangGraph workflows are modeled as directed graphs, where nodes are agents or functions and edges encode how control and data flow between them, often conditioned on the current state. This graph abstraction enables complex behaviors such as retries, revisiting earlier steps, and agents that respond across many turns, rather than forcing everything into a one-shot or strictly linear chain. Moreover, because of the graph architecture, as the number of tools and agents (nodes) increases, the supervisor can use and combine them to perform more robust analysis. This key benefit of the architecture presents an opportunity for research groups in many fields to automate specialized workflows and even combine tools that are currently used in isolation.

Hierarchical agentic frameworks organize multiple AI agents into layered roles, as shown in Figure 1, with a higher-level supervisor agent that decomposes complex goals into structured sub-tasks and delegates them to lower-level specialist agents. These specialist agents use user-defined tools, which they execute and report back to higher-level agents, enabling role specialization across planning, routing, execution, and evaluation. This architecture treats LLMs as reasoning components within a coordinated hierarchy rather than as generic standalone endpoints.[23]

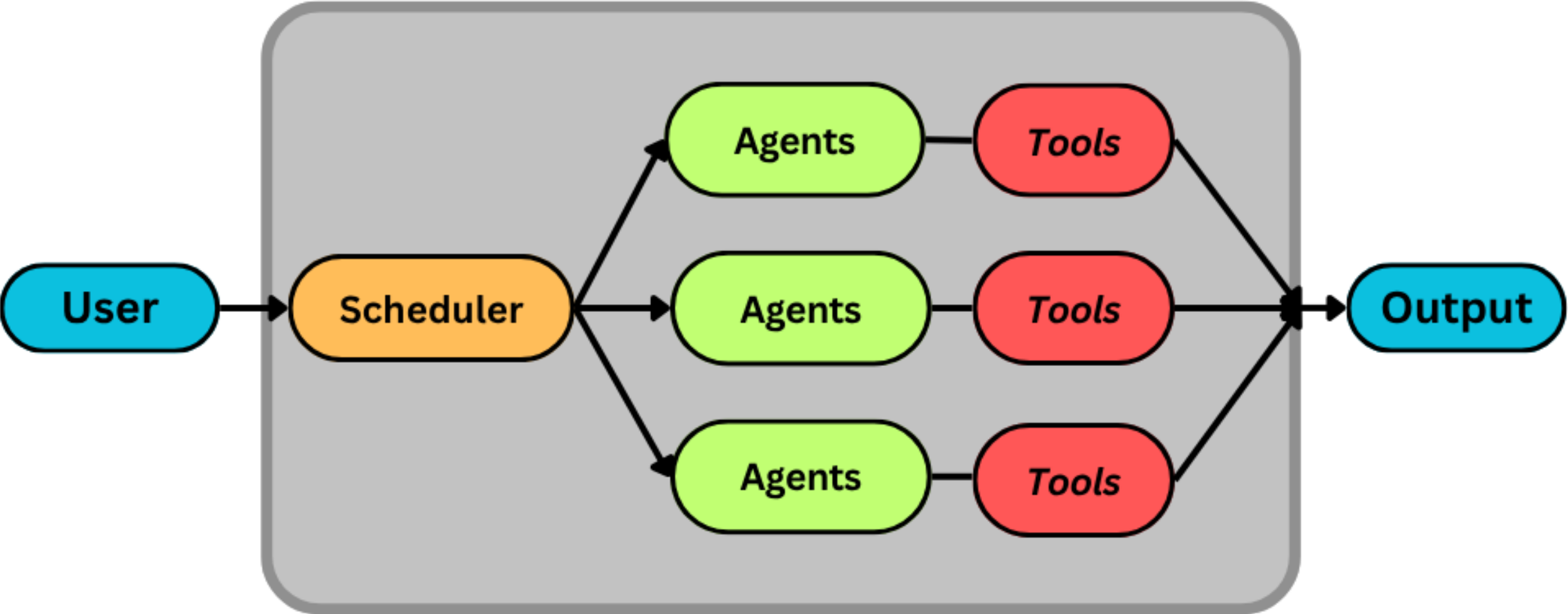

Figure 1. Schematic of the DynaMate[20] hierarchical agentic framework using LangChain. The user interacts with the supervisor, which assigns tasks to specialized agents, equipped with user-defined tools. Solid arrows represent the information flow. Reproduced from Ref. 20 with permission from the Royal Society of Chemistry.

Tools are bound to specific agents and can take many forms: retrievers, APIs, code interpreters, simulators, databases, or actuators, implemented as Python functions. Information flows top-down as high-level intent and task assignments, laterally when a specialist agent coordinates multiple tools for a single sub-task, and bottom-up as structured results and feedback that supervisors aggregate to revise plans, reprioritize tasks, or trigger additional tool calls. Compared to single-agent workflows, hierarchical frameworks provide modularity (components can be swapped or improved independently), specialization (agents and tools optimized for narrow competencies), better error detection through supervisory review, more robust handling of long-horizon tasks through explicit decomposition, and greater scalability, since adding capacity means instantiating new worker agents rather than redesigning a monolithic system. [24]

Recent systems that adopt this pattern include MetaGPT[25], which assigns software-engineering roles to distinct agents coordinated by a shared message pool; CAMEL[26], which pairs a user-proxy agent with a task-solving agent in a role-playing loop; and AutoGen[27], which provides a conversation-driven multi-agent framework in which agents negotiate tool use and verification steps. In scientific computing, analogous designs have been applied to autonomous chemical research (Coscientist[28]), drug discovery[29], and alloy design (AtomAgents[30]), each placing a planner or critic at the top of a hierarchy of specialist workers.

Despite this progress, a critical limitation persists across both general-purpose and domain-specific hierarchical frameworks: the agent architecture is fixed at deployment time. Adding a new tool or agent requires a developer to modify the source code and restart the system. This prevents domain scientists from adapting the system to their specific workflows without programming expertise. DynaMate2 (see Figure 2) addresses this gap by treating the agent pool itself as a dynamic, user-managed resource: tools can be registered from inline code, source files, or natural-language descriptions; agents can be added and removed; and all extensions persist automatically across sessions.

The user (blue block) sends a prompt to the framework, and this is received by the "prompt enhancer" (green block). This agent has knowledge of the system's pool registry and uses this information to improve the prompt and signal which agents, tools, and inputs are needed to achieve the task. When complex tasks are requested, this prompt enhancer creates a plan, and this is then passed to the supervisor agent (purple). The supervisor has access to the tool registry and to all the specialized agents (orange blocks) that have tools that the user added (cyan blocks). Users can add, modify, and remove tools and agents directly from the prompt and can dynamically use them in the following prompt. The prompt enhancer agent, the supervisor, and the specialized agents have access to the persistence layer, which stores all the information and metadata about the user-added tools. In the current implementations, they are saved as Python scripts that the user can read and modify if needed. This behavior enables experts to maintain full control over the scripts and tools used by the agents.

The design goal of DynaMate2 is to reduce the programming effort required for researchers with established Python workflows to integrate their tools into an AI-supervised pipeline. These efforts aim to accelerate the rate of acquisition of scientific knowledge. The general idea is that users can drag and drop their Python functions into the user interface (UI) and dynamically create a custom agentic framework for their specific application. In this work, we use molecular simulations as a test case to demonstrate how scripts that execute a small part of a workflow can be converted into agentic tools that can then be dynamically used to automate the workflow.

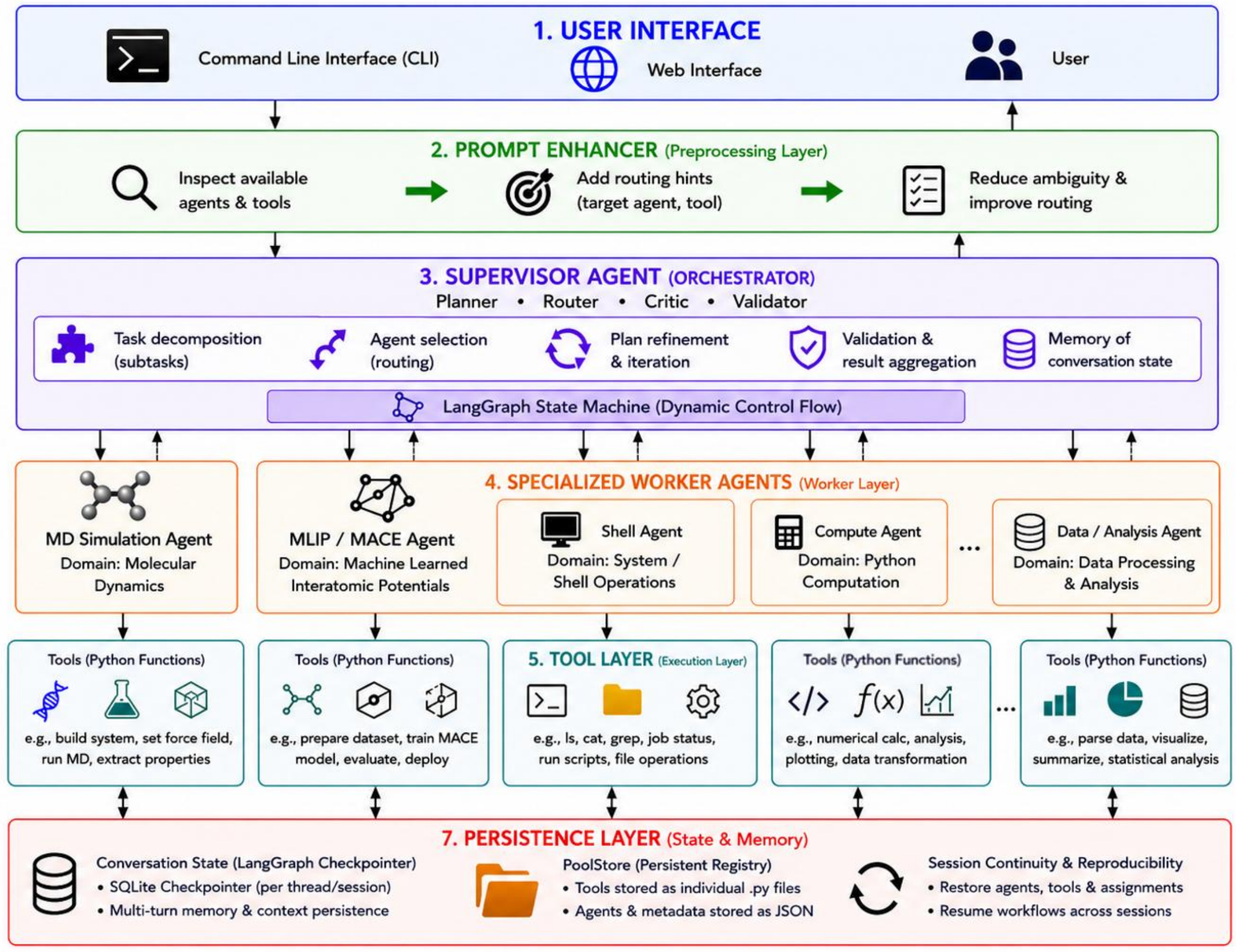

Figure 2. Schematic of the DynaMate2 hierarchical agentic framework. The user interacts with the supervisor, which decomposes a complex prompt into sub-tasks assigned to specialized agents, each equipped with expert-defined tools. Solid arrows represent always-active information flow; dashed lines represent conditional routing.

## 2.1 Core Abstractions and Code Template

DynaMate2 centers on a single Python class, PersistentAgentPoolWithSupervisor, that encapsulates the agent system's full lifecycle. It manages a registry of callable tools, a pool of specialist agents, a LangGraph-compiled supervisor, and a SQLite-backed checkpointer that preserves conversation state across sessions. This design distills the three-tier hierarchy into one coherent object that a researcher can configure, extend, and query without modifying any underlying library code.

The architecture is structured in three layers. The tool layer contains Python callable registered as LangChain StructuredTool objects, each with an auto-generated docstring that the LLM reads when deciding which tool to invoke. The agent layer wraps each specialist in a LangGraph ReAct agent whose effective system prompt is assembled from a user-supplied base description, an automatically generated block listing the currently assigned tools, and a fixed execution rule that prevents the agent from returning control to the supervisor before producing a concrete tool result. The supervisor layer hosts a single routing LLM that receives all user messages, decomposes them into subtasks, and dispatches to the appropriate agent via LangGraph handoff edges.

Tool and agent definitions are persisted to disk by a PoolStore component, which serializes each tool's source code to an individual .py file inside a configurable state directory and records agent metadata to a companion JSON file. At startup, the pool automatically reloads all previously registered tools and agents, so that a long-running research session can be interrupted and resumed without loss of custom extensions. A PromptEnhancer preprocessing module sits between the user interface and the supervisor: it inspects each incoming message, queries the pool for the current roster of agents and tools, and rewrites the message to append a routing hint identifying the most appropriate agent and, when determinable, the specific tool name. This explicit guidance reduces the supervisor's search space and improves routing reliability.[31]

## 2.2 Hierarchical Routing and Execution Control

The supervisor is realized as a LangGraph graph compiled by the langgraph-supervisor library. At initialization, the pool builds a static system prompt that instructs the supervisor on its role and appends a live block listing all currently registered agents, along with the first line of each tool's docstring. This combined prompt is recompiled whenever an agent is added or removed, ensuring the supervisor's knowledge remains consistent with the actual pool state.

Routing proceeds in two stages. First, the PromptEnhancer rewrites the user's message to include a suggested target agent and, when the message unambiguously names a task for which a specific tool exists, the name of that tool. The enhanced message is then passed to the supervisor, which evaluates the routing hint against its agent list and issues a LangGraph handoff to transfer execution to the indicated specialist. If the hint is absent or ambiguous, the supervisor reasons over the available agents and selects the most appropriate one. A configurable recursion limit (default 25 hops) prevents infinite loops when agents return inconclusive results.

Specialist agents operate independently in ReAct loops, as formalized by Yao et al.[32] Upon receiving a task from the supervisor, each agent iterates over its tool list, selects the most appropriate tool, executes it, evaluates the result, and either refines its approach or returns the outcome to the supervisor. The execution rule injected into every agent's system prompt explicitly enumerates its available domain tools and prohibits transferring control back to the supervisor without first calling at least one of them. This constraint closes the most common failure mode observed during development, in which an agent acknowledges the task and immediately returns without performing any work.

State management is handled by a LangGraph SqliteSaver checkpointer keyed on a thread identifier supplied by the caller. Each conversation turn is persisted atomically, allowing multi-turn dialogues in which the supervisor can refer back to results from earlier steps.[33] The web interface and notebook both manage thread identifiers explicitly, creating fresh threads for independent tasks and reusing existing threads to continue iterative workflows.

### 2.3 Base Tools and Default Agents

DynaMate2 ships with two always-present agents and six management tools available to the supervisor from the moment the pool is instantiated. These base components handle system-level operations and runtime extensibility without requiring any user configuration. The shell_agent wraps LangChain's ShellTool and can execute arbitrary shell commands on the host system. The compute_agent is a general-purpose Python execution sandbox for tasks requiring numerical computation or scripting without a persistent shell session.

The six management tools collectively implement the dynamic extensibility that distinguishes DynaMate2 from its predecessor: register_tool_from_code and register_tool_from_file add new callables to the tool registry; assign_tool_to_agent and add_agent_to_pool extend the pool with new capabilities; remove_tool_from_agent and remove_agent_from_pool reduce the pool when capabilities are no longer needed. All six are bound to the supervisor's tool list and can be invoked in response to natural language requests without any manual API calls.

## 3. Tool Registration Protocol

A primary design goal of DynaMate2 is to enable researchers to integrate their existing, validated Python code into the agentic framework with minimal effort. The following protocol describes the complete process for converting an expert-defined function into a registered, LLM-callable tool. This protocol applies whether the function is a standalone script, part of a larger library, or written specifically for DynaMate2. Figure 3 shows the landing page of the DynaMate2 UI, where users find tabs with buttons that pre-fill the prompt box with examples to add functions and run the test discussed later in this document. The predefined prompt can be edited or completely modified for new tasks. The tab for creating tools from scripts includes an option to upload files from your local machine, which triggers a modifiable prompt to convert the script into an agentic tool. Users can access this Gradio[34] user interface (UI) by running the app.py script locally. All files will be saved in this directory, and additional details can be found in the repository storing the code.

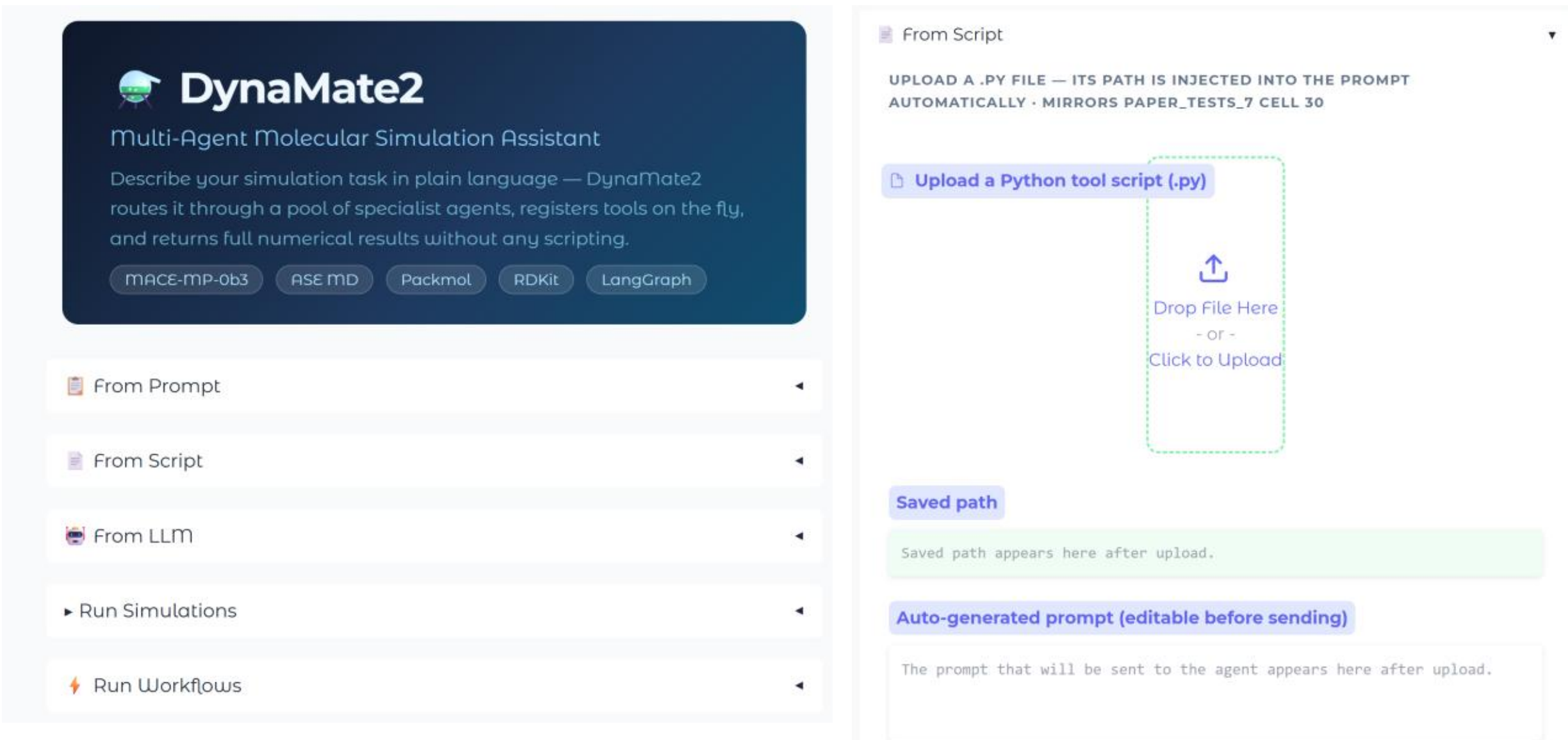

Figure 3. DynaMate2 landing page of the Gradio user-interface (UI). Tabs include buttons for predefined prompts, tools from the prompt, scripts, an LLM description, and options to run simulations and the full MD workflow discussed in this document.

## 3.1 Preparing Your Function

Before registering a tool, the existing function must satisfy two structural requirements that arise from DynaMate2's execution model.

3.1.1 **Place all imports inside the function body.** When a tool is registered via register_tool_from_code, the source string is executed immediately using Python's exec(). Any top-level import statement will therefore run at registration time, not at call time. In a multi-agent environment where heavyweight packages such as MACE[35], ASE[36], or LAMMPS[37] may not be available on every worker, top-level imports will raise an ImportError and prevent registration. Moving all imports into the function body defers them until the moment the tool is actually called, when the appropriate scientific Python stack is guaranteed to be available.

3.1.2 **Write a complete, descriptive docstring.** The tool's docstring is the primary source of information the LLM uses when deciding whether and how to invoke the tool. It should (a) state in plain language what the function does; (b) enumerate all parameters with their types and expected values; and (c) describe the return value. A poorly written or absent docstring will cause the supervisor and specialist agents to misroute or ignore the tool. This is the single most impactful step in the registration process.[38]

## 3.2 Registering the Tool

The UI of DynaMate2 provides three registration pathways, which are presented in Figure 3, each suited to a different situation. When users open these tabs, they will see buttons that, when clicked, fill the prompt input box with predefined prompts that replicate the test case discussed in this document. The user can modify this prompt to use the existing tools, but they can also add new tools and define new prompts to run them. The three main ways to register your existing Python function as an agentic tool are:

3.2.1 **Registration from inline code.** Paste the function source code directly into a natural-language message to the supervisor (e.g., "Register the following tool: [paste code]"). The supervisor routes the request to register_tool_from_code, which executes the source in a controlled namespace, wraps each discovered callable as a LangChain StructuredTool, and adds it to the global registry. This pathway is best for short, self-contained functions developed interactively.

3.2.2 **Registration from a source file.** For functions developed and tested in a standard Python environment and stored in a .py file, instruct the supervisor with a plain-language message (e.g., "Register the tools in my_simulation.py"). The register_tool_from_file function reads the source from disk and delegates to register_tool_from_code. This pathway is recommended for validated, version-controlled production tools, as it allows researchers to develop and test functions in their preferred environment before exposing them to the agent pipeline.

3.2.3 **Registration from a natural-language description.** For new tools that do not yet exist as code, describe the desired function in plain language to the supervisor (e.g., "Create a function that reads an ASE[36] trajectory, extracts per-frame energies and temperature, and produces a two-panel figure"). The tool_manager agent generates a complete Python function from the specification, including internal imports, a descriptive docstring, and the implementation, and automatically registers it, following established LLM code-generation paradigms.[39] The request must not contain a Python def statement, as the PromptEnhancer treats any message containing a def line as a direct code-registration request rather than a code-generation specification.

Once registered, newly added tools appear in pool.list_registered_tools() but are not yet bound to any agent. All three registration pathways persist the tool source code to the state directory via PoolStore, so registered tools survive session restarts.

## 3.3 Creating Specialist Agents and Assigning Tools

3.3.1 **Create a specialist agent.** Issue a natural language instruction to the supervisor (e.g., "Add an agent called md_specialist focused on molecular dynamics simulations"). The add_agent_to_pool function accepts the agent name and a plain-language description of its role and expertise. Internally, it creates a LangGraph ReAct agent, stores the agent's metadata in the pool registry, and recompiles the supervisor graph to include the new agent as a routing option. The supervisor's system prompt is updated automatically.

3.3.2 **Assign tools to the agent.** Issue a follow-up instruction (e.g., "Assign run_nvt_md to md_specialist"). The assign_tool_to_agent function looks up the tool by name, appends it to the agent's extra_tools list, and reconstructs the agent's effective system prompt. The rebuilt prompt lists every assigned tool with the first line of its docstring and re-injects the execution rule requiring the agent to call at least one domain tool before returning control to the supervisor. Both agent metadata and tool assignments are persisted to disk and will be restored automatically in future sessions.

Table 1 below presents a summary of the step-by-step registration protocol for this framework.

Table 1. Summary of the complete registration protocol as a quick reference.

| Step | Action | Details |
|---|---|---|
| **1. Prepare function** | Move all imports inside the function body | Prevents ImportError at registration time |
| **2. Write docstring** | Describe inputs, parameters, and return value in plain language | Docstring is the LLM's only guide to tool selection |
| **3a. Register from code** | Paste the source in a natural-language message to the supervisor | Source string must not have top-level imports |
| **3b. Register from file** | Instruct supervisor: "Register tools in [filename].py" | The file must be accessible at the specified path |
| **3c. Register from description** | Describe the desired function in prose; do not include a def statement | PromptEnhancer interprets def lines as direct code input |
| **4. Create an agent** | Instruct the supervisor to add a specialist agent with a role description | Role description shapes routing and execution behavior |
| **5. Assign tool** | Instruct supervisor to assign [tool_name] to [agent_name] | The agent system prompt is rebuilt automatically |
| **6. Verify** | Call pool.list_registered_tools() and pool.list_agents() | Check the docstring is accurate and the assignment is correct |
| **7. Invoke** | Issue a natural-language prompt to the supervisor | No additional code required |

## 3.4 Verifying and Using Registered Tools

3.4.1 **Verify the registration.** Call pool.list_registered_tools() to confirm the tool appears in the global registry and pool.list_agents() to verify it has been assigned to the correct specialist. Review the auto-generated tool description to ensure the docstring accurately reflects the function's behavior, as this is what the LLM will read.

3.4.2 **Invoke via natural language.** The registered tool is now callable through any natural language prompt directed at the supervisor. The PromptEnhancer will identify the most appropriate agent and tool based on the docstring content, and the supervisor will route the task accordingly. No further API calls or code modifications are required.

# 4. Representative Example: End-to-End MLIP-Based Molecular Dynamics Workflow

To demonstrate the full capabilities of DynaMate2 as an agentic template for expert-defined workflows, we present a complete end-to-end molecular dynamics simulation driven entirely by natural language instructions. The workflow uses the MACE-POLAR[40] family of MLIP foundation models and covers all stages, from model retrieval to trajectory analysis. It was executed in a Jupyter notebook (DynaMate2_tutorial.ipynb) and via the Gradio UI, both connected to a GPU cluster node.

This workflow was constructed incrementally over four tool-building steps (T1–T4), each of which applied the Tool Registration Protocol described in Section 3. The individual registration steps, including input messages, PromptEnhancer outputs, and intermediate outputs, are provided as supplementary information. Here, we focus on the integrated end-to-end execution and the reasoning the supervisor employs to coordinate the workflow.

## 4.1 Tool and Agent Build-Up

Before the end-to-end run, the following tools and agents were registered via the natural language interface, according to the protocol in Section 3. A brief description of each registration step is provided; full source code and details about tool validation are included in the Supplementary Information.

**T1 — Model download tool (inline code registration):** The function download_mace_model, which downloads a named MACE[35] foundation model checkpoint from the public repository to a local models/ directory, was registered by pasting its source code directly into a message to the supervisor. All ASE[36] and MACE imports were placed inside the function body. The function was then assigned to a new specialist agent, mace_md_specialist, created specifically for MLIP-related tasks. (See SI Section 1.1)

**T2 — System construction tools (inline code + agent creation):** Two chemistry tools were registered from inline code in a single message: smiles_to_xyz, which converts a SMILES string to an XYZ coordinate file using RDKit[41], and packmol_build_system, which calls Packmol[42] to construct a periodic simulation box of a molecular system at a target density. Both were assigned to mace_md_specialist. This step demonstrated that multiple tools can be registered and from a single prompt interaction. (See SI Section 1.2)

**T3 — NVT molecular dynamics tool (file-based registration):** The function run_nvt_md, which wraps ASE's[36] Langevin thermostat[43] with the MACE[35] calculator to run constant-temperature MD simulations, was developed and tested independently in a standard Python environment and stored in ASE_NVT_PBC.py. It was registered with the instruction "Register the tools in ASE_NVT_PBC.py" and assigned to mace_md_specialist. This step validates the file-based registration pathway for production-grade, version-controlled scientific tools. (See SI Section 1.3)

**T4 — Trajectory analysis tool (description-based generation):** A trajectory plotting function, plot_nvt_trajectory, was generated entirely by the tool_manager agent from a plain-language description requesting a function that reads an ASE[36] trajectory file, extracts per-frame potential energy, total energy, and temperature, normalizes the energy traces by their time-averages, and produces a two-panel figure. The generated code was registered and assigned to mace_md_specialist automatically. This step demonstrates the most autonomous registration pathway, in which the LLM generates tool code from a specification while still ensuring all domain execution logic is encapsulated in the registered function. (See SI Section 1.4)

## 4.2 End-to-End Execution from a Single Instruction

With all four tools registered and assigned to mace_md_specialist, the complete workflow was executed from a single high-level natural language prompt directed at the supervisor. The prompt requested that DynaMate2 perform the full pipeline: download the MACE-POLAR[40] medium checkpoint, construct a periodic NaCl-water system (i.e. a standard benchmark system for MD force fields[44]) containing 267 water molecules and one $Na^{+}/Cl^{-}$ ion pair in a 20.0 Å cubic box (0.9918 $g/cm^3$), run a 250 ps NVT simulation at 300 K using the downloaded model, and plot the resulting energy and temperature traces. This example is intended to evaluate the orchestration of a multi-step computational workflow; it is not used to draw conclusions about NaCl-water structure, thermodynamics, or MLIP accuracy for solvated ions. Table 2 below presents the user input, prompt enhancer addition, and supervisor output for this test, and Figure 4 presents a visualization of the initial configuration generated and the normalized potential and total energy from the 250 ps simulation. Additional details are available in Section 2 of the SI.

Table 2. Input, prompt enhancer addition, and supervisor output for an end-to-end example using the registered tools for the full workflow involving the generation of an XYZ file of initial configuration, downloading a pre-trained model, running a molecular simulation, and plotting energies and temperatures from the trajectory.

| | |
|---|---|
| Input | Run a complete NVT molecular dynamics study of a NaCl aqueous solution from scratch. Use the MACE-POLAR-M force field and save the raw model to `./models`. Convert the water SMILES (O) to a 3D structure and save it to `./water.xyz`. Convert the NaCl ion pair SMILES ([Na+].[Cl-]) to a 3D structure and save it to `./nacl.xyz`. Build a periodic cubic box of 20.0 Å containing 267 water molecules and 1 NaCl ion pair, and save it to `./nacl_water_box.xyz`. Run a 250 ps NVT simulation at 300 K with a 0.5 fs timestep, saving a trajectory frame every 10 steps, using ASE and save the trajectory to `./nacl_water_nvt.traj`. Finally, plot the potential and total energies and temperature over time and save the figure as a PNG to `nacl_water_nvt_analysis.png`. |
| Prompt Enhancer | - Run a complete NVT molecular dynamics study of a NaCl aqueous solution from scratch using the MACE-POLAR-M force field and save the raw model to `/models`.<br><br>- Convert the water SMILES (O) to a 3D structure and save it to `/water.xyz` using smiles_to_xyz.<br><br>- Convert the NaCl ion pair SMILES ([Na+].[Cl-]) to a 3D structure and save it to `/nacl.xyz` using smiles_to_xyz.<br><br>- Build a periodic cubic box of 20.0 Angstrom containing 267 water molecules and 1 NaCl ion pair, and save it to `/nacl_water_box.xyz` using packmol_build_system.<br><br>- Run a 250 ps (500,000 step) NVT simulation at 300 K with a 0.5 fs timestep, saving a trajectory frame every 10 steps, using ASE and save the trajectory to `/nacl_water_nvt.traj` using run_nvt_md.<br><br>- Finally, plot the potential and total energies and temperature over time and save the figure as a PNG to `nacl_water_nvt_analysis.png` using plot_nvt_trajectory.<br><br>- Use mace_md_specialist — it should use smiles_to_xyz, packmol_build_system, run_nvt_md, and plot_nvt_trajectory to complete the request. All required input files are already present at the provided paths. Call these tools directly without any preliminary file-preparation steps. |
| Output: | The complete NVT molecular dynamics study of the NaCl aqueous solution using the MACE-POLAR-M force field has been completed successfully.<br><br>- The raw model is saved at: /groups/ycolon/Orlando/scratch365/Molec_Mindset/DynaMate2_prod/tutorials/models/MACE-POLAR-M.model<br><br>- Water 3D structure (from SMILES) is saved at: /groups/ycolon/Orlando/scratch365/Molec_Mindset/DynaMate2_prod/tutorials/end_to_end_test/water.xyz<br><br>- NaCl ion pair 3D structure (from SMILES) is saved at: /groups/ycolon/Orlando/scratch365/Molec_Mindset/DynaMate2_prod/tutorials/end_to_end_test/nacl.xyz |
| | - Periodic cubic box with 267 water molecules and 1 NaCl ion pair is saved at:<br>/groups/ycolon/Orlando/scratch365/Molec_Mindset/DynaMate2_prod/tutorials/end_to_end_test/nacl_water_box.xyz<br><br>- The 250 ps NVT simulation trajectory at 300 K is saved at:<br>/groups/ycolon/Orlando/scratch365/Molec_Mindset/DynaMate2_prod/tutorials/end_to_end_test/nacl_water_nvt.traj<br><br>- The plot of potential and total energies and temperature over time is saved as PNG at:<br>/groups/ycolon/Orlando/scratch365/Molec_Mindset/DynaMate2_prod/tutorials/end_to_end_test/nacl_water_nvt_analysis.png<br><br>If you need any further analysis or assistance, please let me know! |

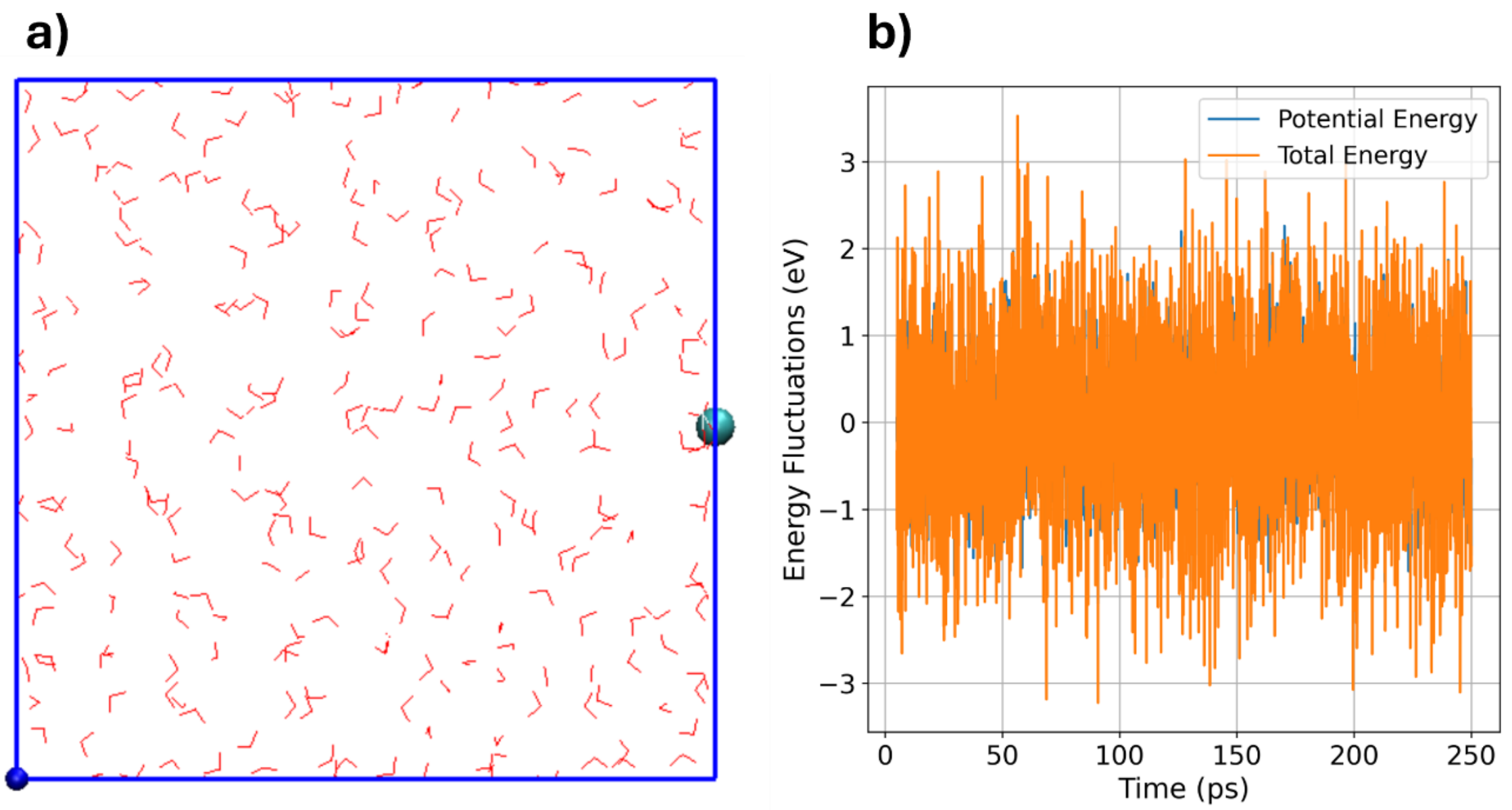

Figure 4. Panel a shows the visualization (generated using VMD[45]) of the initial NaCl-water configuration generated using the packmol tool, containing 267 water molecules, one Na (blue) cation, and one Cl (cyan) anion in a 20.0 Å cubic box. Panel b presents the normalized potential and total energy from the 250 ps simulation.

The supervisor decomposed the high-level instruction into four sequential sub-tasks and routed each to mace_md_specialist. The PromptEnhancer identified the target agent and, in each sub-task, the specific tool to invoke, passing this routing hint to the supervisor. The execution sequence proceeded as follows: (1) download_mace_model was called with the model identifier "MACE-POLAR-1-M", downloading the checkpoint to models/; (2) smiles_to_xyz was called three times—for water (O), $Na^+$ ([Na+]), and $Cl^-$ ([$Cl^-$])—and packmol_build_system was called with the resulting XYZ files, molecular counts, and box dimensions to produce nacl_water_box.xyz; (3) run_nvt_md was called with the configuration file, the downloaded checkpoint, a temperature of 300 K, 500,000 steps, and the output trajectory path nvt_nacl_water.traj; and (4) plot_nvt_trajectory was called with the trajectory path and an output PNG filename, producing the final figure.

The supervisor maintained the thread context across all four sub-tasks, passing file paths and model identifiers generated in earlier steps as inputs to later steps without requiring any user intervention. This demonstrates the core capability that DynaMate2 is designed to provide: a researcher registers their validated domain tools once, and the supervisor handles the coordination, sequencing, and context propagation for multi-step composable workflows.

Compared to the DynaMate workflow for an analogous task, DynaMate2 reduces manual user interventions substantially. DynaMate required a separate prompt for each tool invocation, with the user manually monitoring outputs and deciding whether to proceed or retry, resulting in four or more independent interactions for a workflow of this complexity. DynaMate2 implements the same pipeline within a single session thread from a single instruction.

A practical consideration for end-to-end workflows is thread management. A thread identifier key is used for each session, and the SQLite checkpointer retains the full message history for that thread. Long workflows accumulate context that may influence later routing decisions. It is advisable to use distinct thread identifiers for logically independent sub-tasks and a shared thread for steps that genuinely depend on prior context.[46,47] The notebook examples in the repository illustrate both patterns.

# 5. Conclusion and Outlook

## 5.1 Summary

DynaMate2 addresses a practical barrier that currently limits the adoption of agentic AI frameworks in scientific research: the difficulty of integrating existing, expert-developed code into a supervised multi-agent pipeline. Research groups across computational science have invested years building specialized, validated tools for their specific workflows. DynaMate2 provides the infrastructure and protocol to transform these tools into AI-callable components without requiring researchers to modify framework source code, retrain models, or acquire deep machine learning expertise.

The central design principle is that the LLM is never responsible for generating scientific code, only for routing tasks and coordinating outputs. This is what makes this integration reliable. By keeping all domain logic in expert-defined, validated functions and restricting the LLM to a supervisory and coordinating role, DynaMate2 avoids code-generation errors that could compromise scientific integrity. The LLM's role is precisely defined: decompose the goal, select the registered tool, pass the appropriate inputs, and use the output to plan the next action.

The framework advances the state-of-the-art along three axes relative to its predecessor. First, it replaces flat single-agent routing with a genuine supervisor-specialist hierarchy in which multiple agents collaborate within a single session, unlike single-agent paradigms where one model handles all sub-tasks.[7] Second, it introduces runtime extensibility through the Tool Registration Protocol, enabling tools and agents to be added, modified, and removed through natural language instructions without interrupting an ongoing session. Third, it demonstrates end-to-end MLIP-based molecular dynamics from model download through system construction, NVT simulation, and trajectory analysis, as a concrete example of how multi-step composable scientific workflows can be automated once expert tools are registered.

For the broader scientific community, DynaMate2 offers a template rather than a finished product. A research group working in materials characterization, protein simulation, climate modeling, or any other domain with established Python-based workflows can adopt the framework, register their existing tools following the protocol in Section 3, and immediately benefit from LLM-driven automation and multi-step coordination. The more tools are registered, the more powerful the supervisor's ability to compose them into novel, complex workflows becomes.

## 5.2 Limitations and Open Challenges

Several limitations remain. Agent reliability is the most pressing concern: the current implementation requires carefully engineered execution rules injected into each agent's system prompt to prevent the common failure mode in which a specialist transfers control back to the supervisor without performing any work. This fragility is a direct consequence of the underlying LLM's tendency to interpret ambiguous instructions as opportunities to defer, and it worsens as the number of tools per agent increases. More principled solutions, such as constrained decoding or learned routing policies, are an active area of research.[48]

Tool generation quality is a second limitation. When the system writes a new tool from a natural language description, the generated code may contain subtle errors that go undetected until runtime[49], since the registration step only checks syntactic validity. Automated unit testing of generated tools before registration, or the integration of a critic agent to review generated code, would significantly improve reliability.[50]

The current framework does not support the distributed or parallel execution of multiple specialist agents. All agents run sequentially within a single process, which limits throughput for workflows that could benefit from concurrent tool calls, a known challenge in multi-agent LLM systems.[51] This is a known limitation of the current LangGraph configuration and is a target for future development.

DynaMate2 is intended for supervised local deployment in which the user controls the Python environment and the code being registered. Because the framework can execute shell commands and register functions, it should not be exposed to untrusted users or untrusted code without sandboxing, filesystem restrictions, and explicit permission controls. The current implementation records tool sources and metadata to support inspection and audit, but it does not by itself constitute a security boundary.

## 5.3 Future Directions

Several directions are being pursued for future versions. Parallel agent execution, enabled by LangGraph's native support for concurrent subgraphs, would allow independent subtasks, such as building multiple system configurations simultaneously, to proceed in parallel.[52] Integration with robotic laboratory platforms would extend the framework to closed-loop experimental workflows in which simulated and experimental results inform each other iteratively, as demonstrated in autonomous laboratory platforms. [28,53]

On the machine learning side, replacing heuristic prompt engineering with learned routing policies trained on successful and unsuccessful routing decisions could substantially improve the reliability of agent selection, building on previous work in agent orchestration.[54] A community tool registry, analogous to a package index, would allow research groups to publish and share validated DynaMate2 tools, accelerating adoption across domains beyond molecular dynamics.[55] This is particularly important given the motivation for DynaMate2: the scientific community already has the domain tools and the framework's value grows with each new tool that is registered and shared.

## 6. Data Availability

The DynaMate2 repository includes the core framework, all tools demonstrated in this paper, the DynaMate2_tutorial.ipynb Jupyter notebook that reproduces the representative examples, and a web-based Gradio[34] interface providing a conversational front end to the full agent pool. Detailed documentation and contribution guidelines are provided to encourage community extension of the tool library and agent pool configurations, in line with open-science principles for reproducible and extensible research software.[56] We welcome contributions from researchers in computational chemistry, materials science, and adjacent fields who wish to adapt the framework to their own workflows.

Access at: https://github.com/omendibleba/DynaMate2

## 7. Conflicts of Interest

The authors have no conflicts of interest to disclose.

## 8. Acknowledgements

This work was performed using the computational resources provided by the Notre Dame Center for Research Computing (NDCRC). Y. C. acknowledges funding support from the U.S. Department of Education (award no. P200A210048) and the University of Notre Dame. U. C. acknowledges funding support from the University of Puerto Rico–Mayagüez.

# 9. References

(1) Wang, L.; Ma, C.; Feng, X.; Zhang, Z.; Yang, H.; Zhang, J.; Chen, Z.; Tang, J.; Chen, X.; Lin, Y.; Zhao, W. X.; Wei, Z.; Wen, J. A Survey on Large Language Model Based Autonomous Agents. *Front. Comput. Sci.* **2024**, *18* (6), 186345. https://doi.org/10.1007/s11704-024-40231-1.

(2) M. Bran, A.; Cox, S.; Schilter, O.; Baldassari, C.; White, A. D.; Schwaller, P. Augmenting Large Language Models with Chemistry Tools. *Nat. Mach. Intell.* **2024**, *6* (5), 525–535. https://doi.org/10.1038/s42256-024-00832-8.

(3) McNaughton, A. D.; Sankar Ramalaxmi, G. K.; Kruel, A.; Knutson, C. R.; Varikoti, R. A.; Kumar, N. CACTUS: Chemistry Agent Connecting Tool Usage to Science. *ACS Omega* **2024**, *9* (46), 46563–46573. https://doi.org/10.1021/acsomega.4c08408.

(4) Yang, F.; Evans, J. D. QUASAR: A Universal Autonomous System for Atomistic Simulation and a Benchmark of Its Capabilities. 2026.

(5) Zou, Y.; Cheng, A. H.; Aldossary, A.; Bai, J.; Leong, S. X.; Campos-Gonzalez-Angulo, J. A.; Choi, C.; Ser, C. T.; Tom, G.; Wang, A.; Zhang, Z.; Yakavets, I.; Hao, H.; Crebolder, C.; Bernales, V.; Aspuru-Guzik, A. El Agente: An Autonomous Agent for Quantum Chemistry. *Matter* **2025**, *8* (7). https://doi.org/10.1016/j.matt.2025.102263.

(6) Campbell, Q.; Cox, S.; Medina, J.; Watterson, B.; White, A. D. MDCrow: Automating Molecular Dynamics Workflows with Large Language Models. 2025.

(7) Ramos, M. C.; Collison, C. J.; White, A. D. A Review of Large Language Models and Autonomous Agents in Chemistry. *Chem. Sci.* **2025**, *16* (6), 2514–2572. https://doi.org/10.1039/D4SC03921A.

(8) Zhang, Z.; Yin, A.; Baweja, A.; Bai, J.; Gustin, I.; Bernales, V.; Aspuru-Guzik, A. El Agente Forjador: Task-Driven Agent Generation for Quantum Simulation. 2026.

(9) Zhang, H.; Li, Y.; Huang, W.; Hou, Z.; Song, Y.; Liu, X.; Effaty, F.; Jiang, J.; Wu, S.; Ding, Q.; Takahara, I.; MacGillivray, L. R.; Mizoguchi, T.; Yu, T.; Liao, L.; Luo, Y.; Rong, Y.; Li, J.; Diao, Y.; Ji, H.; Liu, B. Towards Agentic Intelligence for Materials Science. 2026.

(10) Deringer, V. L.; Caro, M. A.; Csányi, G. Machine Learning Interatomic Potentials as Emerging Tools for Materials Science. *Adv. Mater.* **2019**, *31* (46), 1902765. https://doi.org/https://doi.org/10.1002/adma.201902765.

(11) Kang, Y.; Kim, J. ChatMOF: An Artificial Intelligence System for Predicting and Generating Metal-Organic Frameworks Using Large Language Models. *Nat. Commun.* **2024**, *15* (1), 4705. https://doi.org/10.1038/s41467-024-48998-4.

(12) Jiang, G.; Luo, Q. Chemis{TRAG}: Table-Based Retrieval-Augmented Generation for Chemistry Question Answering. 2026.

(13) Zhang, P.; Peng, X.; Han, R.; Chen, T.; Ma, J. Rag2Mol: Structure-Based Drug Design Based on Retrieval Augmented Generation. *Brief. Bioinform.* **2025**, *26* (3), bbaf265. https://doi.org/10.1093/bib/bbaf265.

(14) Kreimeyer, K.; Canzoniero, J. V; Fatteh, M.; Anagnostou, V.; Botsis, T. Using Retrieval-Augmented Generation to Capture Molecularly-Driven Treatment Relationships for Precision Oncology. *Stud. Health Technol. Inform.* **2024**, *316*, 983–987. https://doi.org/10.3233/SHTI240575.

(15) Lewis, P.; Perez, E.; Piktus, A.; Petroni, F.; Karpukhin, V.; Goyal, N.; Küttler, H.; Lewis, M.; Yih, W.; Rocktäschel, T.; Riedel, S.; Kiela, D. Retrieval-Augmented Generation for Knowledge-Intensive NLP Tasks. In *Advances in Neural Information Processing Systems*; Larochelle, H., Ranzato, M., Hadsell, R., Balcan, M. F., Lin, H., Eds.; Curran Associates, Inc., 2020; Vol. 33, pp 9459–9474.

(16) Zhou, J.; Jiang, J.; Han, Z.; Wang, Z.; Gao, X. Streamline Automated Biomedical Discoveries with

Agentic Bioinformatics. *Brief. Bioinform.* **2025**, *26* (5), bbaf505. https://doi.org/10.1093/bib/bbaf505.

(17) Weixin, L.; Yuhui, Z.; Hancheng, C.; Binglu, W.; Yi, D. D.; Xinyu, Y.; Kailas, V.; Siyu, H.; Scott, S. D.; Yian, Y.; A., M. D.; James, Z. Can Large Language Models Provide Useful Feedback on Research Papers? A Large-Scale Empirical Analysis. *NEJM AI* **2024**, *1* (8), AIoa2400196. https://doi.org/10.1056/AIoa2400196.

(18) Pizzi, G.; Cepellotti, A.; Sabatini, R.; Marzari, N.; Kozinsky, B. AiiDA: Automated Interactive Infrastructure and Database for Computational Science. *Comput. Mater. Sci.* **2016**, *111*, 218–230. https://doi.org/https://doi.org/10.1016/j.commatsci.2015.09.013.

(19) Hartung, T. AI, Agentic Models and Lab Automation for Scientific Discovery - the Beginning of ScAInce. *Front. Artif. Intell.* **2025**, *8*, 1649155. https://doi.org/10.3389/frai.2025.1649155.

(20) Mendible-Barreto, O. A.; Díaz-Maldonado, M.; Carmona Esteva, F. J.; Torres, J. E.; Córdova-Figueroa, U. M.; Colón, Y. J. DynaMate: Leveraging AI-Agents for Customized Research Workflows. *Mol. Syst. Des. Eng.* **2025**, *10* (7), 585–598. https://doi.org/10.1039/D5ME00062A.

(21) LangChain. Langchain-Ai/Langgraph. 2025.

(22) Chase, H. LangChain. 2022.

(23) Li, A.; Xie, Y.; Li, S.; Tsung, F.; Ding, B.; Li, Y. Agent-Oriented Planning in Multi-Agent Systems. In *The Thirteenth International Conference on Learning Representations*; 2025.

(24) Guo, T.; Chen, X.; Wang, Y.; Chang, R.; Pei, S.; Chawla, N. V; Wiest, O.; Zhang, X. Large Language Model Based Multi-Agents: A Survey of Progress and Challenges. In *Proceedings of the Thirty-Third International Joint Conference on Artificial Intelligence, {IJCAI-24}*; Larson, K., Ed.; International Joint Conferences on Artificial Intelligence Organization, 2024; pp 8048–8057. https://doi.org/10.24963/ijcai.2024/890.

(25) Hong, S.; Zhuge, M.; Chen, J.; Zheng, X.; Cheng, Y.; Zhang, C.; Wang, J.; Wang, Z.; Yau, S. K. S.; Lin, Z.; Zhou, L.; Ran, C.; Xiao, L.; Wu, C.; Schmidhuber, J. MetaGPT: Meta Programming for A Multi-Agent Collaborative Framework. 2024.

(26) Li, G.; Hammoud, H. A. A. K.; Itani, H.; Khizbullin, D.; Ghanem, B. {CAMEL}: Communicative Agents for "'Mind'" Exploration of Large Language Model Society. In *Thirty-seventh Conference on Neural Information Processing Systems*; 2023.

(27) Wu, Q.; Bansal, G.; Zhang, J.; Wu, Y.; Li, B.; Zhu, E.; Jiang, L.; Zhang, X.; Zhang, S.; Liu, J.; Awadallah, A. H.; White, R. W.; Burger, D.; Wang, C. AutoGen: Enabling Next-Gen LLM Applications via Multi-Agent Conversation. 2023.

(28) Boiko, D. A.; MacKnight, R.; Kline, B.; Gomes, G. Autonomous Chemical Research with Large Language Models. *Nature* **2023**, *624* (7992), 570–578. https://doi.org/10.1038/s41586-023-06792-0.

(29) Seal, S.; Huynh, D. L.; Chelbi, M.; Khosravi, S.; Kumar, A.; Thieme, M.; Wilks, I.; Davies, M.; Mustali, J.; Sun, Y.; Edwards, N.; Boiko, D.; Tyrin, A.; Selinger, D. W.; Parikh, A.; Vijayan, R.; Kasbekar, S.; Reid, D.; Bender, A.; Spjuth, O. AI Agents in Drug Discovery. 2025.

(30) Ghafarollahi, A.; Buehler, M. J. Automating Alloy Design and Discovery with Physics-Aware Multimodal Multiagent AI. *Proc. Natl. Acad. Sci.* **2025**, *122* (4), e2414074122. https://doi.org/10.1073/pnas.2414074122.

(31) Wei, J.; Wang, X.; Schuurmans, D.; Bosma, M.; ichter, brian; Xia, F.; Chi, E.; Le, Q. V; Zhou, D. Chain-of-Thought Prompting Elicits Reasoning in Large Language Models. In *Advances in Neural Information Processing Systems*; Koyejo, S., Mohamed, S., Agarwal, A., Belgrave, D., Cho, K., Oh, A., Eds.; Curran Associates, Inc., 2022; Vol. 35, pp 24824–24837.

(32) Yao, S.; Zhao, J.; Yu, D.; Du, N.; Shafran, I.; Narasimhan, K. R.; Cao, Y. ReAct: Synergizing

Reasoning and Acting in Language Models. In *The Eleventh International Conference on Learning Representations*; 2023.

(33) Park, J. S.; O'Brien, J.; Cai, C. J.; Morris, M. R.; Liang, P.; Bernstein, M. S. Generative Agents: Interactive Simulacra of Human Behavior. In *Proceedings of the 36th Annual ACM Symposium on User Interface Software and Technology*; UIST '23; Association for Computing Machinery: New York, NY, USA, 2023. https://doi.org/10.1145/3586183.3606763.

(34) Abid, A.; Abdalla, A.; Abid, A.; Khan, D.; Alfozan, A.; Zou, J. Gradio: Hassle-Free Sharing and Testing of ML Models in the Wild. *arXiv Prepr. arXiv1906.02569* **2019**.

(35) Batatia, I.; Kovacs, D. P.; Simm, G.; Ortner, C.; Csanyi, G. MACE: Higher Order Equivariant Message Passing Neural Networks for Fast and Accurate Force Fields. In *Advances in Neural Information Processing Systems*; Koyejo, S., Mohamed, S., Agarwal, A., Belgrave, D., Cho, K., Oh, A., Eds.; Curran Associates, Inc., 2022; Vol. 35, pp 11423–11436.

(36) Hjorth Larsen, A.; Jørgen Mortensen, J.; Blomqvist, J.; Castelli, I. E.; Christensen, R.; Dułak, M.; Friis, J.; Groves, M. N.; Hammer, B.; Hargus, C.; Hermes, E. D.; Jennings, P. C.; Bjerre Jensen, P.; Kermode, J.; Kitchin, J. R.; Leonhard Kolsbjerg, E.; Kubal, J.; Kaasbjerg, K.; Lysgaard, S.; Bergmann Maronsson, J.; Maxson, T.; Olsen, T.; Pastewka, L.; Peterson, A.; Rostgaard, C.; Schiøtz, J.; Schütt, O.; Strange, M.; Thygesen, K. S.; Vegge, T.; Vilhelmsen, L.; Walter, M.; Zeng, Z.; Jacobsen, K. W. The Atomic Simulation Environment—a Python Library for Working with Atoms. *J. Phys. Condens. Matter* **2017**, *29* (27), 273002. https://doi.org/10.1088/1361-648X/aa680e.

(37) Thompson, A. P.; Aktulga, H. M.; Berger, R.; Bolintineanu, D. S.; Brown, W. M.; Crozier, P. S.; in 't Veld, P. J.; Kohlmeyer, A.; Moore, S. G.; Nguyen, T. D.; Shan, R.; Stevens, M. J.; Tranchida, J.; Trott, C.; Plimpton, S. J. LAMMPS - a Flexible Simulation Tool for Particle-Based Materials Modeling at the Atomic, Meso, and Continuum Scales. *Comput. Phys. Commun.* **2022**, *271*, 108171. https://doi.org/https://doi.org/10.1016/j.cpc.2021.108171.

(38) Qin, Y.; Liang, S.; Ye, Y.; Zhu, K.; Yan, L.; Lu, Y.; Lin, Y.; Cong, X.; Tang, X.; Qian, B.; Zhao, S.; Hong, L.; Tian, R.; Xie, R.; Zhou, J.; Gerstein, M.; li, dahai; Liu, Z.; Sun, M. ToolLLM: Facilitating Large Language Models to Master 16000+ Real-World APIs. In *International Conference on Learning Representations*; Kim, B., Yue, Y., Chaudhuri, S., Fragkiadaki, K., Khan, M., Sun, Y., Eds.; 2024; Vol. 2024, pp 9695–9717.

(39) Chen, M.; Tworek, J.; Jun, H.; Yuan, Q.; de Oliveira Pinto, H. P.; Kaplan, J.; Edwards, H.; Burda, Y.; Joseph, N.; Brockman, G.; Ray, A.; Puri, R.; Krueger, G.; Petrov, M.; Khlaaf, H.; Sastry, G.; Mishkin, P.; Chan, B.; Gray, S.; Ryder, N.; Pavlov, M.; Power, A.; Kaiser, L.; Bavarian, M.; Winter, C.; Tillet, P.; Such, F. P.; Cummings, D.; Plappert, M.; Chantzis, F.; Barnes, E.; Herbert-Voss, A.; Guss, W. H.; Nichol, A.; Paino, A.; Tezak, N.; Tang, J.; Babuschkin, I.; Balaji, S.; Jain, S.; Saunders, W.; Hesse, C.; Carr, A. N.; Leike, J.; Achiam, J.; Misra, V.; Morikawa, E.; Radford, A.; Knight, M.; Brundage, M.; Murati, M.; Mayer, K.; Welinder, P.; McGrew, B.; Amodei, D.; McCandlish, S.; Sutskever, I.; Zaremba, W. Evaluating Large Language Models Trained on Code. 2021.

(40) Batatia, I.; Baldwin, W. J.; Kuryla, D.; Hart, J.; Kasoar, E.; Elena, A. M.; Moore, H.; Gawkowski, M. J.; Shi, B. X.; Kapil, V.; Kourtis, P.; Magdău, I.-B.; Csányi, G. MACE-POLAR-1: A Polarisable Electrostatic Foundation Model for Molecular Chemistry. 2026.

(41) Landrum, G. RDKit: Open-Source Cheminformatics. 2026.

(42) Martínez, L.; Andrade, R.; Birgin, E. G.; Martínez, J. M. PACKMOL: A Package for Building Initial Configurations for Molecular Dynamics Simulations. *J. Comput. Chem.* **2009**, *30* (13), 2157–2164. https://doi.org/https://doi.org/10.1002/jcc.21224.

(43) Vanden-Eijnden, E.; Ciccotti, G. Second-Order Integrators for Langevin Equations with Holonomic Constraints. *Chem. Phys. Lett.* **2006**, *429* (1), 310–316. https://doi.org/https://doi.org/10.1016/j.cplett.2006.07.086.

(44) Joung, I. S.; Cheatham, T. E. I. I. I. Determination of Alkali and Halide Monovalent Ion Parameters for Use in Explicitly Solvated Biomolecular Simulations. *J. Phys. Chem. B* **2008**, *112* (30), 9020–9041. https://doi.org/10.1021/jp8001614.

(45) Humphrey, W.; Dalke, A.; Schulten, K. VMD -- Visual Molecular Dynamics. *J. Mol. Graph.* **1996**, *14*, 33–38.

(46) Anthropic. Context Windows - Claude API Documentation. 2026.

(47) Liu, N. F.; Lin, K.; Hewitt, J.; Paranjape, A.; Bevilacqua, M.; Petroni, F.; Liang, P. Lost in the Middle: How Language Models Use Long Contexts. *Trans. Assoc. Comput. Linguist.* **2024**, *12*, 157–173. https://doi.org/10.1162/tacl_a_00638.

(48) Kambhampati, S.; Valmeekam, K.; Guan, L.; Verma, M.; Stechly, K.; Bhambri, S.; Saldyt, L.; Murthy, A. LLMs Can't Plan, But Can Help Planning in LLM-Modulo Frameworks. 2024.

(49) Zhang, Z.; Wang, C.; Wang, Y.; Shi, E.; Ma, Y.; Zhong, W.; Chen, J.; Mao, M.; Zheng, Z. LLM Hallucinations in Practical Code Generation: Phenomena, Mechanism, and Mitigation. *Proc. ACM Softw. Eng.* **2025**, *2* (ISSTA). https://doi.org/10.1145/3728894.

(50) Madaan, A.; Tandon, N.; Gupta, P.; Hallinan, S.; Gao, L.; Wiegreffe, S.; Alon, U.; Dziri, N.; Prabhumoye, S.; Yang, Y.; Gupta, S.; Majumder, B. P.; Hermann, K.; Welleck, S.; Yazdanbakhsh, A.; Clark, P. Self-Refine: Iterative Refinement with Self-Feedback. In *Advances in Neural Information Processing Systems*; Oh, A., Naumann, T., Globerson, A., Saenko, K., Hardt, M., Levine, S., Eds.; Curran Associates, Inc., 2023; Vol. 36, pp 46534–46594.

(51) Liu, Z.; Zhang, Y.; Li, P.; Liu, Y.; Yang, D. A Dynamic LLM-Powered Agent Network for Task-Oriented Agent Collaboration. 2024.

(52) Zhuge, M.; Wang, W.; Kirsch, L.; Faccio, F.; Khizbullin, D.; Schmidhuber, J. GPTSwarm: Language Agents as Optimizable Graphs. In *Forty-first International Conference on Machine Learning*; 2024.

(53) Szymanski, N. J.; Rendy, B.; Fei, Y.; Kumar, R. E.; He, T.; Milsted, D.; McDermott, M. J.; Gallant, M.; Cubuk, E. D.; Merchant, A.; Kim, H.; Jain, A.; Bartel, C. J.; Persson, K.; Zeng, Y.; Ceder, G. An Autonomous Laboratory for the Accelerated Synthesis of Inorganic Materials. *Nature* **2023**, *624* (7990), 86–91. https://doi.org/10.1038/s41586-023-06734-w.

(54) Shi, F.; Chen, X.; Misra, K.; Scales, N.; Dohan, D.; Chi, E. H.; Schärli, N.; Zhou, D. Large Language Models Can Be Easily Distracted by Irrelevant Context. In *Proceedings of the 40th International Conference on Machine Learning*; Krause, A., Brunskill, E., Cho, K., Engelhardt, B., Sabato, S., Scarlett, J., Eds.; Proceedings of Machine Learning Research; PMLR, 2023; Vol. 202, pp 31210–31227.

(55) Huber, S. P.; Zoupanos, S.; Uhrin, M.; Talirz, L.; Kahle, L.; Häuselmann, R.; Gresch, D.; Müller, T.; Yakutovich, A. V; Andersen, C. W.; Ramirez, F. F.; Adorf, C. S.; Gargiulo, F.; Kumbhar, S.; Passaro, E.; Johnston, C.; Merkys, A.; Cepellotti, A.; Mounet, N.; Marzari, N.; Kozinsky, B.; Pizzi, G. AiiDA 1.0, a Scalable Computational Infrastructure for Automated Reproducible Workflows and Data Provenance. *Sci. Data* **2020**, *7* (1), 300. https://doi.org/10.1038/s41597-020-00638-4.

(56) Wilkinson, M. D.; Dumontier, M.; et al. The FAIR Guiding Principles for Scientific Data Management and Stewardship. *Sci. Data* **2016**, *3* (1), 160018. https://doi.org/10.1038/sdata.2016.18.